\documentclass[letterpaper,twocolumn,10pt]{article}
\usepackage{usenix-2020-09}

\usepackage{tikz}
\usepackage{amsmath}

\usepackage{filecontents}

\usepackage[flushleft]{threeparttable}

\usepackage{halloweenmath}
\usepackage{subcaption}
\usepackage{amsthm}
\usepackage{amsfonts}
\usepackage[ruled,linesnumbered]{algorithm2e}
\newcommand*{\defeq}{\stackrel{\text{def}}{=}}
\let\oldnl\nl
\newcommand{\nonl}{\renewcommand{\nl}{\let\nl\oldnl}}
\SetKwInput{KwData}{Input}
\SetKwInput{KwResult}{Output}
\DontPrintSemicolon

\usepackage{xcolor}
\usepackage{url}
\usepackage{booktabs}
\newtheorem{proposition}{Proposition}

\usepackage{mathtools}

\definecolor{codegreen}{rgb}{0,0.6,0}
\definecolor{codegray}{rgb}{0.5,0.5,0.5}
\definecolor{codepurple}{rgb}{0.58,0,0.82}
\definecolor{backcolour}{rgb}{0.95,0.95,0.92}
\usepackage{listings}
\lstdefinestyle{mystyle}{
    backgroundcolor=\color{backcolour},   
    commentstyle=\color{codegreen},
    keywordstyle=\color{magenta},
    numberstyle=\tiny\color{codegray},
    stringstyle=\color{codepurple},
    basicstyle=\ttfamily\footnotesize,
    breakatwhitespace=false,         
    breaklines=true,                 
    captionpos=b,                    
    keepspaces=true,                 
    numbers=left,                    
    numbersep=5pt,                  
    showspaces=false,                
    showstringspaces=false,
    showtabs=false,                  
    tabsize=2
}

\theoremstyle{definition}

\usepackage[utf8]{inputenc}

\begin{document}

\date{}

\title{High-Performance Caching of Homomorphic Encryption for Cloud Databases}

\author{
{\rm Dongfang Zhao}\\
University of Washington\\
{\rm \texttt{dzhao@uw.edu}}
} 

\maketitle

\begin{abstract}
While homomorphic encryption (HE) has garnered significant research interest in cloud-based outsourced databases due to its algebraic properties over ciphertexts, the computational overhead associated with HE has hindered its widespread adoption in production database systems.
Recently, a caching technique called Radix-based additive caching of homomorphic encryption (Rache) was proposed in SIGMOD'23. 
The primary objective of this paper is to address the performance overhead resulting from the expensive randomization process in Rache. To achieve this, we propose a novel encryption algorithm called $ASEnc$, which replaces the computationally intensive full scan of radixes with the caching of a polynomial number of radix-powers during an offline stage. This design significantly reduces the performance impact caused by randomization.
Furthermore, this paper aims to extend Rache's capabilities to support floating-point numbers. To accomplish this, we introduce a new encryption algorithm named $FSEnc$, leveraging efficient constant multiplication available in state-of-the-art fully homomorphic encryption (FHE) schemes. Notably, $FSEnc$ offers the flexibility to cache the coefficients instead of the radixes themselves, which may result in a large number of cached ciphertexts. However, we manage this efficiently by streaming the dynamically cached ciphertexts through a vector of circular buffers.
We demonstrate that both encryption algorithms guarantee semantic security (IND-CPA). To validate their performance, we implement both algorithms as loadable functions in MySQL 8.0 and deploy the system prototype on a 96-core server hosted in the Chameleon Cloud. Experimental results showcase that $ASEnc$ outperforms Rache by 2.3--3.3$\times$, while $FSEnc$ surpasses the state-of-the-art floating-point FHE CKKS by 1.8--5.6$\times$.
\end{abstract}

\section{Introduction}

\subsection{Background}

With the increasing deployment of outsourced databases on public cloud platforms, the security of cloud databases has become a major concern, particularly for applications that handle sensitive data in fields such as public health~\cite{tkanwal_cluster21}, bioinformatics~\cite{xzhu_tdsc21}, and financial services~\cite{wkuan_clsr18}. While encryption schemes like AES~\cite{aes}, RSA~\cite{rsa}, and ECC~\cite{ecc} are available, their use in a typical cloud computing scenario can undermine the purpose and benefits of cloud computing:
Encrypting sensitive data, uploading it to the cloud, downloading portions of data to local storage, and decrypting ciphertexts for computation or analysis results in the cloud serving solely as remote backup storage, devoid of its computing functionality.

One viable solution to address the security concerns in cloud-based outsourced databases is the application of \textit{homomorphic encryption} (HE) schemes. These schemes enable cloud servers to perform computations directly on encrypted records. If an HE scheme supports both addition and multiplication operations, it theoretically allows the composition and computation of any finite functions on the ciphertexts. However, fully homomorphic encryption (FHE)~\cite{ckks,bgv,bfv,bv11}, which supports both addition and multiplication over ciphertexts, tends to introduce a significant performance overhead, making it impractical for time-sensitive applications, such as high-performance scientific computing~\cite{otawose_sigmod23,aalmamun_sc21,xwang_aaai20}. As a result, many production systems opt for HE schemes that support a single algebraic operation (either addition or multiplication) on ciphertexts, commonly known as partial homomorphic encryption (PHE) schemes, e.g.~\cite{ppail_eurocrypt99,symmetria_vldb20}.

To enhance the encryption rate of HE in time-sensitive applications like OLTP in outsourced databases on the public cloud, Rache~\cite{otawose_sigmod23} was proposed. Rache achieves faster encryption by caching specific radix-powers, enabling the construction of ciphertexts for arbitrary plaintexts using the cached radix-powers. Compared to vanilla HE schemes, Rache achieves a speedup of over 10$\times$. However, the randomization of ciphertexts becomes a new performance bottleneck, as the construction process requires a full scan of the cached radixes in addition to the radix arithmetic. Furthermore, Rache is limited to caching integers, leaving the question of caching floating-point ciphertexts in homomorphic encryption unanswered.

\subsection{Motivation and Objectives}

To illustrate the performance bottleneck of state-of-the-art HE schemes in the video analysis ecosystem~\cite{bhaynes_sigmod21,mdaum_icde21}, we provide a real-life example. In scenarios where existing data is updated or new data arrives, Symmetria~\cite{symmetria_vldb20}, the best option available, can only (re-)encrypt the entire dataset. However, Symmetria can encrypt 32-bit random integers at a rate of 3 Mbps, significantly lower than the typical network bandwidth that ranges from tens of Mbps to Gbps. Even with the 10$\times$ performance boost provided by Rache~\cite{otawose_sigmod23}, the performance gap remains significant in practice. Therefore, for data-intensive applications, the encryption subsystem, rather than the I/O subsystem, becomes the performance bottleneck.

This paper has two main goals.

The first goal is to reduce the performance overhead resulting from the expensive randomization process in Rache. To achieve this, we propose a new encryption algorithm called $ASEnc$, which replaces the costly full scan of radixes with the caching of a polynomial number of radix-powers during an offline stage. Although $ASEnc$ involves more cached ciphertexts than Rache, we restrict the number of cached ciphertexts to be a negligible function relative to the security parameter, ensuring semantic security. Additionally, we leverage available CPU cores to parallelize the caching process of the polynomial number of ciphertexts.

The second goal is to extend Rache to support floating-point numbers. We introduce a new encryption algorithm called $FSEnc$, which capitalizes on the efficient constant multiplication available in modern fully HE schemes. In $FSEnc$, we cache the coefficients instead of the radixes themselves. The only drawback of this approach is the potential increase in the number of cached ciphertexts. However, we address this issue by streaming the dynamically cached ciphertexts through a vector of circular buffers.

Our main results are as follows.
From a theoretical perspective, we demonstrate the semantic security of both encryption algorithms. From a practical standpoint, we implement both algorithms as loadable functions in MySQL 8.0 using C++. Furthermore, we deploy a system prototype on a 96-core server hosted in the Chameleon Cloud~\cite{katek_atc20}. Experimental results show that $ASEnc$ outperforms Rache by 2.3--3.3$\times$, while $FSEnc$ outperforms the state-of-the-art floating-point FHE scheme CKKS by 1.8--5.6$\times$.

\subsection{Contributions}

In summary, this paper makes the following contributions:

\begin{itemize}

\item We introduce a new encryption algorithm, $ASEnc$, that enables efficient caching of integer polynomials with inexpensive randomization, outperforming the state-of-the-art caching scheme Rache~\cite{otawose_sigmod23}. (Section \ref{sec:asenc_scheme})

\item We propose a new encryption algorithm, $FSEnc$, specifically designed for streaming cached floating-point ciphertexts in fully homomorphic encryption schemes,
e.g., CKKS~\cite{ckks}. (Section \ref{sec:fsenc_scheme})

\item We provide theoretical proof of the correctness and security (IND-CPA) of both encryption algorithms. (Sections \ref{sec:asenc_analysis} and \ref{sec:fsenc_analysis})

\item We implement the proposed algorithms and integrate them into a production database system through loadable functions. These functions are deployed on the popular Chameleon Cloud~\cite{katek_atc20}. (Section \ref{sec:streche})

\item We conduct an extensive evaluation of the outsourced database system prototype using seven benchmarks. The experimental results demonstrate promising performance, achieving a speedup of up to 5.6$\times$ compared to the state-of-the-art. (Section \ref{sec:eval})
    
\end{itemize}
\section{Preliminaries}
\label{sec:prelim}

\subsection{Homomorphic Encryption}
The notion of \textit{homomorphism} originates from the study of algebraic groups~\cite{fraleigh_book03}, which are algebraic structures defined over nonempty sets. Formally, a group $G$ over a set $S$ is represented as a tuple $(G, \oplus)$, where $\oplus$ is a binary operator satisfying four axioms or properties, expressed as first-order logical formulas:
(i) For all $g, h \in S$, $g \oplus h \in S$.
(ii) There exists a unique element $u \in S$ such that for all $g \in S$, $(g \oplus u = g)$ and $(u \oplus g = g)$.
(iii) For every $g \in S$, there exists an element $h \in S$ such that $(g \oplus h = u)$ and $(h \oplus g = u)$. This element $h$ is often denoted as $-g$.
(iv) For all $g, h, j \in S$, $(g \oplus h) \oplus j = g \oplus (h \oplus j)$.
If we have another group $(H, \otimes)$ and a function $\varphi: G \rightarrow H$ such that for all $g_1, g_2 \in G$, $\varphi(g_1) \otimes \varphi(g_2) = \varphi(g_1 \oplus g_2)$, then we call the function $\varphi$ a homomorphism from $G$ to $H$. In other words, the function $\varphi$ preserves the group operation between the elements of $G$ when mapped to the corresponding elements in $H$.

\textit{Homomorphic encryption} (HE) is a specific type of encryption where certain operations between operands can be performed directly on the ciphertexts.
For example, if an HE scheme $he(\cdot)$ is additive,
then the plaintexts with $+$ operations can be translated into a homomorphic addition $\oplus$ on the ciphertexts.
Formally, if $a$ and $b$ are plaintexts, then the following holds:
\[
dec(he(a) \oplus he(b)) = a + b,
\]
where $dec$ denotes the decryption algorithm.
As a concrete example, let $he(x) = 2^x$, and we temporarily release the security requirement of $he(\cdot)$.
In this case, $he(a+b) = 2^{a+b} = 2^a \times 2^b = he(a) \times he(b)$,
meaning that $\oplus$ is the arithmetic multiplication $\times$.

An HE scheme that supports addition is said to be \textit{additive}.
Popular additive HE schemes include Paillier~\cite{ppail_eurocrypt99},
which is an asymmetric scheme where a pair of public and private keys are used for encryption and decryption.
An HE scheme that supports multiplication is said to be \textit{multiplicative}.
Symmetria~\cite{symmetria_vldb20} is a recent scheme proposed in the database community,
which is multiplicative using a distinct scheme from the one for addition.
Other well-known multiplicative HE schemes include RSA~\cite{rsa} and ElGamal~\cite{elgamal_tit85}.
Similarly, a multiplicative HE scheme guarantees the following equality,
\[
dec (he(a) \otimes he(b)) = a \times b,
\]
where $\otimes$ denotes the homomorphic multiplication over the ciphertexts.

An HE scheme that supports both addition and multiplication is called a \textit{fully HE (FHE) scheme}.
This requirement should not be confused with specific addition and multiplication parameters, such as Symmetria~\cite{symmetria_vldb20} and NTRU~\cite{ntru}.
That is, the addition and multiplication must be supported homomorphically under the same scheme $he(\cdot)$:
\[\displaystyle
\begin{cases}
    dec( he(a) \oplus he(b) ) = a + b \\
    dec( he(a) \otimes he(b) ) = a \times b.
\end{cases}
\]
It turned out to be extremely hard to construct FHE schemes until Gentry~\cite{cgentry_stoc09} demonstrated such a scheme using lattice ideals.
Although lattice has been extensively studied in cryptography,
the combination of lattices and ring ideals is somewhat less explored;
nonetheless, Gentry showed that it is possible to construct an FHE scheme although the cost to maintain the multiplicative homomorphism is prohibitively high even with the so-called bootstrapping optimization,
which essentially applies decryption for every single multiplication operation between ciphertexts.

In the second generation of FHE schemes,
e.g., \cite{bv11,bgv,bfv,ckks},
the encryption efficiency has been greatly improved partially due to the removal of ideal lattices;
rather the new series of FHE schemes are based on the learning with error (LWE) or its variant ring learning with error (RLWE),
which have been proven to be as secure as hard lattice problems (e.g., through quantum or classical reduction).
The good news is that schemes that are built upon LWE or RLWE are significantly more efficient than the first-generation schemes.
However, there is still a wide gap between the state-of-the-art FHE cryptosystems and the desired performance.
Popular open-source libraries of FHE schemes include IBM HElib~\cite{helib}, 
Microsoft SEAL~\cite{sealcrypto},
and others.

\subsection{Threat Model and Provable Security}

We assume the outsourced database servers are semi-honest.
When developing a new encryption scheme, it is crucial to assess its security level, preferably in a provable manner. One widely accepted approach, striking a balance between efficiency and security, is to consider the scenario where the adversary can launch a chosen-plaintext attack (CPA). In a CPA, the adversary can obtain the ciphertext of any arbitrarily chosen plaintext.
In practice, it is assumed that the adversary has limitations on the number of plaintext-ciphertext pairs they can obtain. Specifically, the adversary should only be able to access a polynomial number of such pairs, and their computational resources should be bounded by polynomial time. This assumption reflects the notion that adversaries should not have unlimited computational power or the ability to gather an excessive amount of information.
By making these assumptions, security analysis can be conducted under realistic constraints, allowing for a comprehensive evaluation of the encryption scheme's resilience against chosen-plaintext attacks. The goal is to design an encryption scheme that remains secure even when faced with an adversary who can selectively choose plaintexts and obtain their corresponding ciphertexts within the given limitations.

Ideally, even if the adversary $\mathcal{A}$ can obtain those extra pieces of information, $\mathcal{A}$ should not be able to make a \textit{distinguishably} better decision for the plaintexts than a random guess. This concept is quantified using the notion of a \textit{negligible function}.
A function $\mu(\cdot)$ is considered negligible if, for all polynomials $poly(n)$, the inequality $\mu(n) < \frac{1}{poly(n)}$ holds for sufficiently large values of $n$. 
In other words, the function $\mu(\cdot)$ decreases faster than the reciprocal of any polynomial as $n$ grows. Negligible functions are used to express the level of advantage an adversary can gain over a random guess, and they play a crucial role in the analysis of security proofs for cryptographic schemes.

\section{Polynomial Caching of Integers in Additive Homomorphic Encryption}
\label{sec:asenc}

\subsection{Overview}

\paragraph{Intuition}
We start with the streaming version of Rache~\cite{otawose_sigmod23}.
The idea is straightforward:
Instead of caching only one single ciphertext for each radix,
we now allow the client to compute,
in an offline stage,
multiple ciphertexts for each radix.
Indeed, doing so would increase the precomputing time.
However, such precomputation can be either carried out in a completely offline stage or a streaming fashion.
This offline or streaming mechanism is also widely used in secure multiparty computation,
where a multiplication triplet is precomputed~\cite{dbeaver_crypto91}.

\paragraph{Challenges}
The challenges for trading precomputation of ciphertexts for real-time encryption are twofold.
Firstly, the number of precomputed ciphertexts should be carefully quantified. 
Ideally, the precomputed ciphertexts should allow the database system to sufficiently generate additional randomness to the real-time encryption.
Secondly, 
we need to ensure that the additionally cached ciphertexts do not introduce security vulnerabilities.
For instance, if the original FHE scheme is provably secure (e.g., IND-CPA),
the scheme with the proposed encryption algorithms should be also provably secure.

\paragraph{Notations}
We denote the Rache scheme~\cite{otawose_sigmod23} as a 5-tuple of algorithms $Rache(KeyGen, Enc, Dec, EvalAdd, FastEnc)$.
Generally speaking, all but the $Rache.FastEnc$ algorithm \textit{inherit} from the base homomorphic encryption (HE) scheme,
such as Paillier~\cite{ppail_eurocrypt99} and Symmetria~\cite{symmetria_vldb20}.
As the name suggests, $Rache.Enc$ denotes the original encryption algorithm of the base HE scheme,
while $Rache.FastEnc$ denotes the cached version of the encryption algorithm leveraging the pool of precomputed ciphertexts.
We denote a plaintext message $m \in \mathbb{Z}$,
i.e., we limit our discussion to the message space of integers,
$\mathcal{M} = \mathbb{Z}$.
Let $r$ denote the radix of Rache,
then any $m$ can be written as a polynomial of $r$,
i.e., $m \in \mathbb{Z}^n_r[r]$:
\[
m = c_n \cdot r^n + c_{n-1} \cdot r^{n-1} + \dots + c_1 \cdot r + c_0,
\]
where $c_i \in [0, r)$ denotes the coefficient of the $i$-th power of radix, $i \in [0, n]$.
In general, for an additive HE scheme,
we can further expand each term from the product form into a sequence of additions:
\[
c_i \cdot r^i = \underbrace{r^i + \dots + r^i}_{c_i \textit{ terms}}.
\]
Therefore, any plaintext message $m$ can be written as a sequence of additions and all the summands can be fetched from the pool of cached ciphertexts in $Rache$.
However, in the proposed scheme, 
each $r^i$ can be represented by distinct values.
We use $d_i$ to denote the number of distinct encrypted radixes in the proposed scheme.
We use the superscript to denote the index of the distinct ciphertexts;
e.g., $d_i^{j}$ indicates the coefficient of the $j$-th ciphertext of $r^i$.
According to this definition, 
we have $d_i = \{0, 1\}$ in Rache;
in this sense, Rache can be thought of as a special case of the proposed scheme.

\subsection{Schemes}
\label{sec:asenc_scheme}

\paragraph{Review of \textit{Rache.FastEnc}}
To fully understand the encryption algorithm of the proposed scheme,
namely AStreche,
we provide a quick review of the encryption algorithm of Rache.
We refer interested readers to the original paper~\cite{otawose_sigmod23} for greater detail.
A ciphertext $ctxt$ in Rache can be computed as follows:
\[
ctxt \defeq FastEnc(m) \gets \left(\bigoplus_{i=1}^n c_i \cdot Enc(r^i)\right) \oplus CZero(n),
\]
where $\oplus$ denotes the addition operation defined over the ciphertexts and $CZero(n)$ is the encryption of zeros that are constructed from a Bernoulli process with $n$ experiments and probability $p = 0.5$.
Specifically, $\forall i \in (1,n]$, 
each $r^i$ has a 50\% chance to be added to the ciphertext and if so, $r^{i-1}$ is subtracted $r$ times.
Therefore, the expectation of the number of ciphertext additions for computing $CZero$ is
\[
\sum_{i=2}^{n-1} p \cdot (r + 1) = \frac{(n-2)(n+1)(r+1)}{4},
\]
which is quadratic in $n$.
By contrast, the non-random computation involves the following number of ciphertext additions:
\[
\sum_{i=1}^n |c_i| \le nr,
\]
which is linear in $n$.
Therefore, the major cost of $Rache.FastEnc$ lies in the randomization rather than the ``meaningful'' computation.
However, $CZero$ is essential to the provable security of Rache because the former ensures that the probability for a PPT adversary to break Rache is negligible under the IND-CPA security model.

\paragraph{Scheme of AStreche}
We formalize the proposed scheme as a 5-tuple
\[
AStreche(KeyGen, Enc, Dec, EvalAdd, ASEnc).
\]
Just like Rache, the first four algorithms in AStreche inherit from the base additive HE scheme.
The \textit{KeyGen} takes in a security parameter $1^\lambda$ and generates a secret key $sk$ and optionally a public key $pk$ in case of an asymmetric setting.
The \textit{Enc} algorithm takes in the plaintext message $m$ and the $sk$/$pk$ to return a randomized ciphertext $ctxt$.
The \textit{Enc} algorithm must be a randomized algorithm,
denoted by the $\gets$ symbol $ctxt \gets Enc_{pk}(m)$,
which is required by the CPA threat model.
The \textit{Dec} algorithm, however,
is a deterministic algorithm taking in a (valid) ciphertext to its original plaintext:
$Dec: \{0, 1\}^* \times \mathcal{K} \to \{\bot\} \cup \mathcal{M}$,
where $\bot$ implies an invalid input $ctxt$ of $Dec$ (that is, there is no plaintext $ptxt$ such that $Enc(ptxt) = ctxt$),
$\mathcal{K}$ denotes the key space,
and $\mathcal{M}$ denotes the plaintext massage space.
The $EvalAdd$ algorithm is a deterministic algorithm that takes in two ciphertexts and returns a new ciphertext:
$ctxt_3 = EvalAdd_{pk}(ctxt_1, ctxt_2)$.
The fifth algorithm corresponds to the $FastEnc$ algorithm in Rache;
we elaborate on the $ASEnc$ algorithm in the following section.

\paragraph{Description of \textit{AStreche.ASEnc}}
The algorithm is depicted in Alg.~\ref{alg:asenc}.
The algorithm begins by initializing the radix pool (lines 1--6), which is the set of precomputed ciphertexts corresponding to different powers of $r$. This initialization can be parallelized by employing multiple cores, as the cached ciphertexts are independent. The ciphertext of zero is also precomputed as a starting point (line 7).
To construct the randomized ciphertext $ctxt$ for a given plaintext $ptxt$, the algorithm iterates until the plaintext becomes zero (lines 8--20). At each iteration, it checks the coefficient of the polynomial expansion of $ptxt$ with respect to $r$. Two cases are considered:
(i) If the coefficient is zero, indicating that the current $r$-power is not present in the plaintext (lines 11--13), the $ctxt$ is incremented and decremented by the same randomly-sampled $r$-power. This ensures that the randomization process does not reveal any information about the plaintext.
(ii) Otherwise, if the coefficient is non-zero, the $ctxt$ is incremented by the lower-positioned $r$-power a number of times equal to the coefficient (lines 15--17). This step incorporates the contribution of the $r$-power to the ciphertext.
After each iteration, the plaintext is updated by dividing it by $r$ (line 19), reducing the power of $r$ in the subsequent iteration.
Finally, the algorithm returns the resulting randomized ciphertext $ctxt$.
Please note that the algorithm assumes the availability of an additive HE scheme and the proper implementation of the homomorphic encryption operations.

\begin{algorithm}[!h]
\SetAlgorithmName{Algorithm}{}{}
\caption{ASEnc: Additive Streche Encryption}\label{alg:asenc}
\KwData{
A base homomorphic encryption scheme,
quadruple $HE(KeyGen, Enc, Dec, EvalAdd\;\oplus)$;
A plaintext message $ptxt$;
Radix $r$ for the polynomial expansion of $ptxt$;
The maximal exponent $n$ for radixes;
The data type of ciphertext $Ct$;
}
\KwResult{
A ciphertext \textit{ctxt} such that $Dec_{sk}(ctxt) = ptxt$,
where secret key $sk$ is generated by $KeyGen$
}
 
\nonl \;
\nonl // Initialization (offline) \;
$array<vector<Ct>, n+1> rp$ // radix pool $rp$ \;
\For{$i = 0; i \le n; i++$}{
    \For{$j = 0; j < 2r; j++$}{
        $rp[i].push\_back(Enc_{pk}(i))$ \;
    }
}
$ctxt \gets Enc_{pk}(0)$ \;
 
\nonl \;
\nonl // Encryption (online) \;
\For{$i = 1; i \le n; i++$}{
    $rem \coloneqq ptxt \textit{ \% } (r^i)$ \;
    $idx0 \gets [0, 2r)$ // Randomization \;  
    \eIf{0 = rem}{
        $idx1 \gets [0, 2r)$ // Rehabilitation \;    
        $ctxt \coloneqq ctxt \oplus rp[i][idx0] \ominus rp[i][idx1]$\;
    }
    {
        \For{$j = 0; j < rem; j++$}{
            $ctxt \coloneqq ctxt \oplus rp[i-1][idx0]$\;
        }
    }
    $\displaystyle
    ptxt \coloneqq \left\lfloor \frac{ptxt}{r^i} \right\rfloor$\;
}

\end{algorithm}

\subsection{Analysis}
\label{sec:asenc_analysis}

\paragraph{Correctness}
We assume that $n$ is sufficiently large such that $r^n \ge ptxt$.
The correctness of Alg.~\ref{alg:asenc} can be verified by a series of algebraic derivations from the input to the output;
however, we will provide a simple proof by induction.
That is, we will consider $ptxt = 0$ as the base case and then demonstrate that $ptxt + 1 = Dec(ASEnc(ptxt + 1))$ assuming $ptxt = Dec(ASEnc(ptxt))$.
For the sake of clarity, we will drop the key parameter when referring to the encryption and decryption algorithms in the following.

\begin{itemize}
    \item Base $ptxt = 0$.
    In this case, $rem$ is always 0,
    and only Lines 12--13 are invoked.
    In each iteration, say $i \in [1,n]$, 
    $ctxt$ is added by a random ciphertext of $r^i$ (say, $ctxt_{i}^0$) and then subtracted by another random ciphertext of $r^i$ (say, $ctxt_{i}^1$). 
    Note that both $ctxt_i^0$ and $ctxt_i^1$ are randomly sampled.
    Therefore, the eventual ciphertext can be written as
    \[\displaystyle
    \begin{split}
        ctxt &= ASEnc(0)\\ 
            &= Enc(0) \oplus \left( \bigoplus_{i=1}^n \left( ctxt_i^0 \ominus ctxt_i^1 \right) \right)\\
            &= Enc(0) \oplus \left( \bigoplus_{i=1}^n \left( Enc(r^i) \ominus Enc(r^i) \right) \right),
    \end{split}
    \]
    which means
    \[\displaystyle
    \begin{split}
            & Dec(ctxt) \\
            =\; & Dec\left( Enc(0) \oplus \left( \bigoplus_{i=1}^n \left( Enc(r^i) \ominus Enc(r^i) \right) \right) \right)\\
            =\; & Dec(Enc(0)) + \sum_{i=1}^n  Dec(Enc(r^i)) - Dec(Enc(r^i))\\
            =\; & 0 + \sum_{i=1}^n \left( r^i - r^i \right) = 0 + \sum_{i=1}^n 0 = 0, 
    \end{split}
    \]
    as desired.
    
    \item Induction: $ptxt = Dec(ASEnc(ptxt))$.
    Let $ctxt'$ denote the ciphertext returned by $ASEnc(ptxt')$.
    There are two subcases when $ptxt' \coloneqq ptxt + 1$:
    (i) No digit is carried and (ii) Digits are carried.
    The first subcase (i) is straightforward because the only change lies in $rem$ that is incremented in Lines 14--18:
    \[\displaystyle
    \begin{split}
        Dec(ctxt') &= Dec(Enc(ptxt + 1)) \\
            & = Dec(ctxt \oplus Enc(1)) \\
            & = Dec(ctxt) + Dec(Enc(1))\\
            & = ptxt + 1.
    \end{split}
    \]
    The subcase (ii) occurs when 
    \[
        ptxt \equiv -1 \text{ (mod r)},
    \]
    which implies that $rem = r-1$ when $i = 1$ in Line 9.
    This means that the new remainder $rem'$ would be zero because 
    \[
    rem + 1 = r-1+1 \equiv 0 \text{ (mod r)}.
    \]
    In the meantime, the iteration of $i = 2$ either increments $rem$ by 1 or carries the 1 to the higher $r$-power.
    The latter case is simply analogous to the case of $i=1$,
    so we will deal only with the former case.
    When $i = 1$ and $rem \coloneqq rem + 1$,
    Lines 15--17 imply that 
    \[
        ctxt \coloneqq ctxt \oplus Enc(r),
    \]
    due to the definition and calculation in Lines 16 and 4.
    As a result, the new ciphertext equals
    \[\displaystyle
    \begin{split}
        ctxt' &= ctxt \oplus Enc(r) \ominus Enc(r-1),
    \end{split}
    \]
    and therefore the new plaintext is 
    \[\displaystyle
    \begin{split}
        & ptxt' \\
        =\; & Dec(ctxt')\\ 
        =\; & Dec(ASEnc(ptxt) \oplus Enc(r) \ominus Enc(r-1))\\
        =\; & Dec(ASEnc(ptxt)) + Dec(Enc(r)) - Dec(Enc(r-1))\\
        =\; & ptxt + r - (r-1)\\
        =\; & ptxt + 1,
    \end{split}
    \]
    as desired.
\end{itemize}

\paragraph{Complexity}
The time complexity of Alg.~\ref{alg:asenc} is as follows.
The offline stage takes $\mathcal{O}(nr)$ time,
which does not count toward the execution time of applications.
The online stage takes $\mathcal{O}(nr)$ time as well,
which is linear to the maximal exponent $n$.
In practice, $r$ is usually a small integer;
for example, in Rache~\cite{otawose_sigmod23},
authors show that $r \in [2, 6]$ should suffice for most real-world applications.
On the other hand, $n$ can be thought of as on par with the security parameter---usually in the order of hundreds or thousands. 
Discrete readers would realize the asymptotic time complexity of Alg.~\ref{alg:asenc} and that of Rache are the same;
however, the concrete time costs are different.
In Rache, the overall number of ciphertext summations is $nr + 3n$ because every zero ciphertext for randomization is composed of two additions of $r^{i-1}$ and one subtraction of $r^{i}$,
$i \in [1, n]$;
by contrast, the overall number of ciphertext summations for the randomization of Alg.~\ref{alg:asenc} is $n\cdot \max(r, 2)$ because of Lines 11-18.
The savings during the online stage is not free,
as the offline stage incurs the somewhat expensive $Enc$ operations $nr$ times.

\paragraph{Security}
To demonstrate that the proposed $ASEnc$ algorithm maintains semantic security, we aim to show that the introduction of the radix pool does not weaken the security level of the base HE scheme. This proof follows the typical approach of reduction, where we construct a reduction from a hypothetical adversary who can break the security of the proposed scheme.
We assume that breaking the semantic security of the base HE scheme is a hard problem, believed to be infeasible under the CPA (chosen-plaintext attack) threat model. We proceed with a proof by contradiction, assuming that breaking the proposed $ASEnc$ scheme is an easier problem than breaking the base HE scheme.
In the formal proof (detailed in the technical report), we establish a reduction from breaking the semantic security of the base HE scheme to breaking the $ASEnc$ scheme. This reduction demonstrates that if there exists an adversary capable of breaking $ASEnc$, we can use this adversary to break the semantic security of the base HE scheme.
By assuming the hardness of breaking the base HE scheme and showing that breaking $ASEnc$ leads to breaking the base HE scheme, we conclude that $ASEnc$ is semantically secure as long as the base HE scheme is.
For a comprehensive understanding of the proof, please refer to the detailed formal proof provided in the technical report.
We provide the sketch of the proof in the following.

\begin{proposition}
    AStreche is CPA-secure as long as the base homomorphic encryption scheme 
    \[
    HE(KeyGen, Enc, Dec, EvalAdd)
    \]
    is CPA-secure.
\end{proposition}
\begin{proof}[Proof (Sketch)]
Suppose AStreche is not CPA-secure. Let $\mathcal{A}$ be an adversary who can launch an arbitrary probabilistic polynomial-time (PPT) algorithm to distinguish the ciphertexts of two plaintexts $m_0$ and $m_1$ with a non-negligible probability. Here, we abuse the notion of $n$ to represent the length of the plaintexts, assuming $r = 2$. However, it is worth noting that the following analysis can be extended to $r > 2$.
By convention, we denote the experiment where $\mathcal{A}$ guesses the correct $m_b$, where $b \gets {0, 1}$, as $AStreche_\mathcal{A}^{cpa}$. In this context, the following inequality holds:
    \[\displaystyle
        Pr\left[ AStreche_\mathcal{A}^{cpa}(n) = 1 \right] > \frac{1}{2} + \frac{1}{poly(n)}.
    \]
We introduce the function $oracle$ to represent the black-box query service available to the adversary $\mathcal{A}$ for retrieving ciphertexts. It is important to note that our assumption limits $\mathcal{A}$ to query only a polynomial number of ciphertexts.
Now, let's consider two cases:
(i) In the first case, $\mathcal{A}$ does not access any of the ciphertexts in the radix pool, which we denote as $oracle(\mathcal{A}) \cap rp = \emptyset$.
(ii) In the second case, $\mathcal{A}$ somehow manages to retrieve some ciphertexts that are cached in $rp$, implying that $oracle(\mathcal{A}) \cap rp \neq \emptyset$.
In case (i), the scheme $AStreche$ effectively reduces to the base homomorphic encryption scheme $HE$. We can represent this symbolically as follows:
    \[\displaystyle
    \begin{split}
        &Pr\left[ AStreche_\mathcal{A}^{cpa}(n) = 1\;|\; oracle(\mathcal{A}) \cap rp = \emptyset \right] \\
        =\; &Pr\left[ HE_\mathcal{A}^{cpa}(n) = 1\;|\; oracle(\mathcal{A}) \cap rp = \emptyset \right].
    \end{split}
    \]    
In case (ii), we are aware that the number of ciphertexts cached in $rp$ is bounded by $2r(n+1)$ (refer to Lines 2-6 in Algorithm \ref{alg:asenc}). This bound is polynomial in $n$. As a result, we can establish the following inequality:
    \[\displaystyle
    \begin{split}
        &Pr\left[ AStreche_\mathcal{A}^{cpa}(n) = 1\;|\; oracle(\mathcal{A}) \cap rp \not= \emptyset \right] \\
        <\; &Pr\left[ HE_\mathcal{A}^{cpa}(n) = 1\;|\; oracle(\mathcal{A}) \cap rp \not= \emptyset \right] + \frac{poly(n)}{2^n}.
    \end{split}
    \]
    Combining all of the above three equations,
    we have
    \[\displaystyle
    \begin{split}
        &Pr\left[ HE_\mathcal{A}^{cpa}(n) = 1 \right] \\
        =\; &Pr\left[ HE_\mathcal{A}^{cpa}(n) = 1\;|\; oracle(\mathcal{A}) \cap rp = \emptyset \right]\\
        &+ Pr\left[ HE_\mathcal{A}^{cpa}(n) = 1\;|\; oracle(\mathcal{A}) \cap rp \not= \emptyset \right] \\
        >\; &Pr\left[ HE_\mathcal{A}^{cpa}(n) = 1\;|\; oracle(\mathcal{A}) \cap rp = \emptyset \right]\\
        &+ Pr\left[ HE_\mathcal{A}^{cpa}(n) = 1\;|\; oracle(\mathcal{A}) \cap rp \not= \emptyset \right] - \frac{poly(n)}{2^n}\\
        =\; &Pr\left[ AStreche_\mathcal{A}^{cpa}(n) = 1 \right] - \frac{poly(n)}{2^n}\\
        >\; &\frac{1}{2} + \frac{1}{poly(n)} - \frac{poly(n)}{2^n}\\
        =\; &\frac{1}{2} + f(n),
    \end{split}
    \]
    where $f(n) = \frac{1}{poly(n)} - \frac{poly(n)}{2^n}$.
    But $f(n)$ is not a negligible function because the second term is negligible to the first term and the first term is not negligible in $n$.
    Therefore, we have shown that the base scheme $HE$ is not semantically secure,
    which contracts our assumption and completes the proof.
\end{proof}

\section{Streamed Caching of Floating-point Numbers in Fully Homomorphic Encryption}
\label{sec:fsenc}

\subsection{Overview}

We switch our focus from additive homomorphic encryption to fully homomorphic encryption (FHE) in this section.
We start by articulating the technical challenges of encrypting floating-point numbers with FHE.

\paragraph{Challenges}
When the plaintext $ptxt$ is a floating number,
the most straightforward way to encrypt it is to scale $ptxt$ up by a factor,
e.g., $10^f$, $f \in \mathbb{Z}^+$.
The scale factor $f$ can be selected depending on the precision of the floating numbers in the application.
For example, if the values are stored with three decimal digits,
$f=3$ is sufficient to convert all floating numbers into integers and the method discussed in the previous section can be applied. 
While this scaling technique seems plausible,
at least in theory,
it is not deemed a practical solution because the multiplication between ciphertexts in FHE renders the noises to grow at an exponential rate,
which implies that the scaling factor $f$ contributes to such a fast-growing noise exponentially.
In summary, the key challenge lies in how to efficiently encrypt floating-point numbers without ciphertext inflation.

\paragraph{Notations}
When working with FHE schemes,
users are provided with two binary algebraic operations, 
$\oplus$ and $\otimes$,
to evaluate the summation and production between two ciphertexts.
Furthermore, an FHE scheme often supports \textit{efficient} constant multiplication,
denoted by $\odot$,
over the ciphertext space: 
\[
Dec_{sk}(c \odot ctxt) = c \cdot Dec_{sk}(Enc_{pk}(ptxt)) = c\cdot ptxt,
\]
where $c$ is a constant number (either an integer or a floating number).
For example, the radix $r$ value discussed in the previous section can be used as a constant factor in FHE.
It should be noted that the above constant multiplication can be implemented by a sequence of ciphertext additions with $\oplus$;
for example, Paillier~\cite{ppail_eurocrypt99} can multiply the ciphertext $c$ times to achieve $c \odot ctxt$.
However, this procedure in additive HE schemes is significantly more expensive than FHE (e.g., BGV~\cite{bgv}, CKKS~\cite{ckks}) because the ciphertexts of the latter manipulate vectors and scalars whose constant multiplication can be directly implemented on scalars and entries of the vector.

\paragraph{Intuition}
When dealing with a floating-point number $ptxt$,
we should not be restricted to considering the $ptxt$ as a whole;
rather, we can view a floating-point number as a pair of an integer and a decimal value strictly smaller than one.
Notionally, we have
\[
ptxt = ptxt_i + ptxt_d,
\]
in which $ptxt_i$ denotes the integer part of $ptxt$ and $ptxt_d$ denotes the decimal part of $ptxt$.
For $ptxt_i$, we can apply Alg.~\ref{alg:asenc} because an integer can always be expanded into a polynomial of radix,
denoted by $r_i$.
However, we will encrypt $ptxt_i$ with a more efficient operation through constant multiplication,
which is also the key operation for encrypting $ptxt_d$.
We cannot directly apply Alg.~\ref{alg:asenc} for $ptxt_d$ because we could not reconstruct a decimal using integers:
we cannot apply $\oplus$ a fraction of times.
Here comes the key difference between constant multiplication and sequence of addition:
an expression $r_d \odot ctxt$ in FHE does allow $r_d$ to be a floating-point number.
One of the most straightforward picks of $r_d$ is the reciprocal of 10,
i.e., a $ptxt_d < 1$ with $n_d$ digits can be written as
\[
ptxt_d = c_{-1} \cdot r_d + \dots + c_{-i} \cdot r_d^i + \dots + c_{-n_d} \cdot r_d^{n_d},
\]
where $c_{-i}$ denotes the coefficient of the $i$-th $r_d$-power.
The indices are negative to underline that those coefficients are for decimals rather than that for integers (which are denoted by positive indices).
It should be noted that by definition $c_{-i} \in \left[0, \frac{1}{r_d}\right)$.

\subsection{Schemes}
\label{sec:fsenc_scheme}

\paragraph{Base FHE Scheme}
In this work, we assume that a conventional FHE scheme can be represented by a 6-tuple 
\[\displaystyle
FHE\left(KeyGen, Enc, Dec, Add\;\oplus, Mul\;\otimes, ConstMul\;\odot\right),
\]
where the last three algorithms take in two arguments and return a new ciphertext.
Note that the domains of $\otimes$ and $\odot$ are different:
the former is $\mathcal{C} \times \mathcal{C}$ and the latter is $\mathcal{M} \times \mathcal{C}$,
where $\mathcal{C}$ denotes the ciphertext space and $\mathcal{M}$ denotes the plaintext space.

\paragraph{Streche FHE Scheme for Floating-point Numbers}
Building upon an FHE scheme, the proposed full Streche (FStreche) scheme is a 7-tuple:
\[\displaystyle
FStreche(KenGen, Enc, Dec, \oplus, \otimes, \odot, FSEnc),
\]
where the last algorithm $FSEnc$ is the accelerated version of encryption and the other algorithms inherit from the base FHE scheme.
$FSEnc$ is semantically the same as $Enc$ but the former exhibits significant speedup over the latter if stream caching is enabled,
as we will elaborate on in the next section.

\paragraph{Encryption Algorithm}
The encryption algorithm is presented in Alg.~\ref{alg:fsenc}.
The key idea of the $FSEnc$ algorithm is to introduce streamed caching that leverages the $\odot$ operation provided by the base FHE scheme as a black box.
The cached ciphertexts are the possible coefficients of integer and decimal radixes,
as shown in Lines 1--2.
Note that this is in contrast to the cached ciphertexts in Alg.~\ref{alg:asenc},
which precomputes the radix-powers in the pool.
The reason for this change is that $FSEnc$ applies the constant multiplication between the plaintext and the ciphertext spaces,
and therefore we do have the choice to cache either factor around $\odot$.
Evidently, it is more flexible to cache the coefficients than the radix-powers because the former exhibits a finer granularity (i.e., linear vs. exponential).
Lines 3--10 initialize the radix pools for both the integer part ($ptxt_i$, Lines 4--6) and the decimal part ($ptxt_d$, Lines 7--9).
Line 11 initializes the returned ciphertext $ctxt$ as the encryption of zero,
which can be efficiently retrieved from the first element (i.e., the $front$) of the circular buffer corresponding to the 0-indexed array in the radix pool of integer ciphertexts,
i.e., $rp_i[0][0]$.
Every cached ciphertext can be used only once;
therefore, Line 12 invalidates (i.e., pops) the front of the circular buffer and Line 13 implies that the background process (i.e., a daemon process) streams a newly generated ciphertext to the end of the circular buffer.
Lines 14--21 encrypt the integer part $ptxt_i$.
Just as Alg.~\ref{alg:asenc}, 
$ptxt_i$ is encrypted on a radix basis (Lines 15 and 20) and the ciphertext $ctxt$ is updated by a product between the radix power and the coefficient (Line 17).
Line 16 uniformly samples an index $salt$ in the radix pool and compensate $salt$ with appropriate offset to return the correct coefficient in Line 17. 
We use $rp^+$ to denote that if the index of $rp$ is negative,
then the element with the absolute value of the negative index is returned and the element is negated.
That is, if $idx < 0$,
then $rp^+[idx][0] = -1 \odot rp[-idx][0]$;
otherwise, $rp^+[idx][0] = rp[idx][0]$.
Lines 18--19 work analogously as Lines 12--13:
each cached ciphertext in the pool can be used once after which a background process feeds a new ciphertext to the pool.
Lines 22--29 encrypt the decimal part $ptxt_d$.
As before, Lines 23 and 28 imply that the encryption is carried out on a radix basis,
Line 25 updates the ciphertext with the product of the radix-power and the coefficient (with randomness introduced in Line 24),
and Lines 26--27 update the radix pool.

\begin{algorithm}[!h]
\SetAlgorithmName{Algorithm}{}{}
\caption{FSEnc: Full Streche Encryption}\label{alg:fsenc}
\KwData{
A base fully homomorphic encryption scheme,
$FHE(KeyGen, Enc, Dec, \oplus, \otimes, \odot)$;
A plaintext message $ptxt$;
Radix $r_i$ for the integer part of $ptxt$;
Radix $r_d$ for the decimal part of $ptxt$;
The maximal exponent $n_i$ for $r_i$;
The maximal exponent $n_d$ for $r_d$;
The data type of ciphertext $Ct$;
The length of the circular buffer $L$;
}
\KwResult{
A ciphertext \textit{ctxt} s.t. $Dec_{sk}(ctxt) = ptxt$;
secret key $sk$ generated by $KeyGen$
}
 
\nonl \;
\nonl // Initialization (offline) \;
$array<circular\_buffer<Ct>, r_i+1> rp_i$ \;
$array<circular\_buffer<Ct>, \frac{1}{r_d}> rp_d$ \;
\For{$j = 0; j < L; j++$}{
    \For{$idx = 0; idx \le r_i; idx++$}{
        $rp_i[idx].push\_back(Enc_{pk}(idx))$ \;
    }
    \For{$idx = 1; idx \le \frac{1}{r_d}; idx++$}{
        $rp_d[idx-1].push\_back(Enc_{pk}(idx))$ \;
    }    
}
$ctxt \coloneqq rp_i[0][0]$ \;
$rp_i[0].pop\_front()$\;
$rp_i[0].push\_back(Enc_{pk}(0))$ // Streaming daemon\;
 
\nonl \;
\nonl // Encryption (online) \;
\For{$idx = 1; idx \le n_i; idx++$}{
    $rem \coloneqq ptxt_i \textit{ \% } r_i$ \;
    $salt \gets [0, r_i)$ // Randomization\;
    $ctxt \coloneqq r_i^{idx} \odot \left( rp_i[salt][0] \oplus rp^+_i[rem-salt] \right) \oplus ctxt$\;
    $rp_i[salt].pop\_front()$\;
    $rp_i[salt].push\_back(Enc_{pk}(salt))$ // Streaming\;
    $ptxt_i \coloneqq \left\lfloor \frac{ptxt_i}{r_i} \right\rfloor$\;
}
\For{$idx = 1; idx \le n_d; idx++$}{
    $rem \coloneqq \left\lfloor \frac{ptxt_d}{r_d} \right\rfloor$ \;
    $salt \gets [0, r_d-1)$ // Randomization\;
    $ctxt \coloneqq r_d^{idx} \odot \left( rp_d[salt][0] \oplus rp^+_d[rem-salt][0] \right) \oplus ctxt$\;
    $rp_d[salt].pop\_front()$\;
    $rp_d[salt].push\_back(Enc(salt))$ // Streaming\;
    $ptxt_d \coloneqq ptxt_d - rem$\;
}
\end{algorithm}

\subsection{Analysis}
\label{sec:fsenc_analysis}

\paragraph{Correctness}
Unfortunately, we could not use induction to prove the correctness of Alg.~\ref{alg:fsenc} because floating-point numbers are not countable.
That is, even if we could derive $Dec(FSEnc(ptxt + 1)) = ptxt + 1$ based on the assumption of $Dec(FSEnc(ptxt)) = ptxt$,
there are still infinitely many values falling in the range $(ptxt, ptxt+1)$. 
We will demonstrate $Dec(FSEnc(ptxt)) = ptxt$ using direct computation,
as follows.
By definition, we have 
\[\displaystyle
\begin{split}
Dec(FSEnc(ptxt)) &= Dec(FSEnc(ptxt_i + ptxt_d))\\
    &= Dec(FSEnc(ptxt_i) \oplus FSEnc(ptxt_d))\\
    &= Dec(ctxt_i \oplus ctxt_d),
\end{split}
\]
where $ctxt_i$ corresponds to $ctxt$ on Line 21 of Alg.~\ref{alg:fsenc} and $ctxt_d$ denotes the change of $ctxt$ between Line 21 and Line 29 of Alg.~\ref{alg:fsenc}.
Looking at Lines 14--21, we know
\[\displaystyle
\begin{split}
ctxt_i &= \bigoplus_{idx=1}^{n_i} \left( r_i^{idx} \odot Enc\left(ptxt_i \;\%\; r^{idx}_i \right) \right)\\
    &= \bigoplus_{idx=1}^{n_i} \left( Enc\left( r_i^{idx} \cdot \left( ptxt_i \;\%\; r^{idx}_i \right) \right) \right)\\
    &= Enc\left( \sum_{idx = 1}^{n_i} r_i^{idx} \cdot \left( ptxt_i \;\%\; r^{idx}_i \right) \right)\\
    &= Enc\left(  ptxt_i \right).
\end{split}
\]
Similarly, Lines 22--29 lead to the following calculation:
\[\displaystyle
\begin{split}
ctxt_d &= \bigoplus_{idx=1}^{n_d} \left( r_d^{idx} \odot Enc\left( \left\lfloor \frac{ptxt_d}{r^{idx}_d} \right\rfloor \right) \right)\\
    &= \bigoplus_{idx=1}^{n_d} \left( Enc\left( r_d^{idx} \cdot \left\lfloor \frac{ptxt_d}{r^{idx}_d} \right\rfloor \right) \right)\\
    &= Enc\left( \sum_{idx = 1}^{n_d} r_d^{idx} \cdot \left\lfloor \frac{ptxt_d}{r^{idx}_d} \right\rfloor \right)\\
    &= Enc\left(  ptxt_d \right).
\end{split}
\]
It follows that
\[\displaystyle
\begin{split}
Dec(FSEnc(ptxt)) &= Dec(Enc(ptxt_i) \oplus Enc(ptxt_d))\\
&= Dec(Enc(ptxt_i)) + Dec(Enc(ptxt_d))\\
&= ptxt_i + ptxt_d\\
&= ptxt,
\end{split}
\]
as desired.

\paragraph{Complexity}
To somewhat simplify the notation,
let $n = n_i + n_d$.
It is evident that the offline stage takes $\mathcal{O}(n \cdot L)$ if we assume the cost of $Enc$ takes a constant period.
The online stage takes a linear time in $n$ because Lines 14--27 essentially encrypt $ptxt$ on each $r_i$-power and $r_d$-power term.
Although the asymptotic cost of the online phase is $\mathcal{O}$,
it should be clear that the concrete cost involves constant multiplication and ciphertext addition,
both of which are considered constant operations.
We will have a better understanding of these operations in the evaluation section.

\paragraph{Security}
As $ASEnc$ of Alg.~\ref{alg:asenc},
we must show that $FSEnc$ is CPA-secure.
While a similar proof to that of Alg.~\ref{alg:fsenc} can be provided,
i.e., reducing the problem of breaking the ciphertext generated by $Enc$ to the problem of breaking a ciphertext generated by $FSEnc$,
we will briefly sketch the CPA security using the well-known one-time pad (OTP) scheme,
which is provably secure.
This boils down to showing that the ciphertext generated by $FSEnc$ must be indistinguishable from a random element in the ciphertext space
Recall that the base FHE is presumably CPA-secure.
Therefore, each ciphertext cached in the radix pool is indistinguishable from a random string in $\mathcal{C}$,
the ciphertext space.
Based on Lines 5, 8, 11, 17, and 25 of Alg.~\ref{alg:fsenc},
all ciphertexts generated by $FSEnc$ comprise at least one $Enc$,
whose output is indistinguishable from a random string per our assumption.
In fact, due to Lines 12-13, 18--19, and 26--27,
such ciphertexts from the radix pool are,
informally,
analogous to the ciphertexts of an OTP;
this is so because the ciphertexts are never used more than one time because of Lines 12, 18, and 26 in Alg.~\ref{alg:fsenc}.
Therefore, the ciphertexts generated by $FSEnc$ should look sufficiently random---computationally indistinguishable from true randomness.

\section{System Implementation in Cloud Databases}
\label{sec:streche}

We have successfully implemented both AStreche and FStreche on the MySQL 8.0 database instance hosted at the Chameleon Cloud~\cite{katek_atc20} using loadable functions. The loadable functions are implemented in C++, while the client is implemented in Python using the \textit{pymysql} library. Additionally, we have developed various configuration files and Bash scripts to support the functionality.
As of the current writing, the entire code base comprises 3,413 lines of code.
The source code is accessible through a shared image titled \textit{Streche} on the Chameleon Cloud.

\subsection{User-Defined Functions}

MySQL offers three methods for implementing user-defined functions (UDFs):

SQL-based UDFs: Users can define custom functions using SQL statements. This approach is the most convenient way to extend MySQL's functionality. However, it has limitations in terms of flexibility, especially for complex computations based on number theory, as required in modern cryptography. Additionally, performance tuning of SQL-based UDFs is not well-supported and can be challenging.

Source code modification: Since MySQL is open-source, users have the option to modify the source code of MySQL to implement their UDFs. Modifying the source code allows for fine-tuning system performance. However, this approach is more error-prone, time-consuming, and requires recompiling and testing the entire system. Moreover, source code modifications are difficult to port or scale to existing deployment.

Loadable functions: Loadable functions offer a good trade-off between flexibility and performance. They can be seamlessly integrated into MySQL without the need to recompile the entire system, while still delivering comparable performance to source code modifications. Loadable functions are shared objects that can be separately compiled from the source C++ files. This approach provides developers with nearly the same level of flexibility as source code modifications. We have implemented the proposed Streche schemes as loadable functions, leveraging this approach.

\subsection{Circular Buffers}
Circular buffers are widely employed in various operating system scenarios, such as CPU slicing, as well as in database systems like buffer pools. Their ring-like topology enables efficient retrieval of the head element and convenient addition of elements to the tail, with the head and tail positions being dynamically updated. These features are highly desirable for managing the cached ciphertexts of coefficients in the FSEnc scheme, considering that the coefficient volume exceeds that of the radix-powers cached by ASEnc. Utilizing a circular buffer effectively helps restrict memory consumption.

In our implementation, we utilize the Boost library in C++ to implement circular buffers. Further system details can be found in the Evaluation section. Essentially, our implementation maintains a vector of circular buffers, with each buffer corresponding to a specific radix power coefficient as illustrated in Algorithm \ref{alg:fsenc}.

\subsection{Parallel Caching}
Offline caching incurs significant costs for both ASEnc and FSEnc. While this caching does not affect real-time application execution, reducing the performance overhead is still desirable. The precomputation of certain elements' ciphertexts, such as radix-powers or their coefficients, can be embarrassingly parallel. Thus, we can leverage thread-level parallelization within loadable functions to improve performance.

This is another compelling reason to implement Streche in loadable functions. Introducing thread-level parallelism through libraries like OpenMP is relatively straightforward, which is nontrivial when working with SQL-level UDFs or modifying the entire MySQL source code base.

In our implementation of loadable functions, we utilize OpenMP pragmas to parallelize the caching process across available threads. However, employing multiple threads also introduces overhead, such as context switching. Additionally, each thread may not be allocated sufficient resources to efficiently process its assigned task, especially if it is a hyperthread rather than a physical core. These factors can lead to a counterintuitive effect where more threads result in reduced performance. We will observe and analyze this phenomenon in the Evaluation section.

\section{Evaluation}
\label{sec:eval}

\subsection{Experimental Setup}

\subsubsection{Test Bed}

For our system prototype deployment, we utilize the Chameleon Cloud infrastructure, specifically reserving the c07-08 instance hosted by the Texas Advanced Computing Center (TACC). The c07-08 instance offers the following specifications.
CPUs: Two Intel Gold 6240R CPUs running at a frequency of 2.40GHz, providing a total of 96 threads.
Storage: A 480 GB Micron SATA SSD.
RAM: 192 GB of memory.
These hardware specifications provide the necessary resources for running our system prototype efficiently.

In our c07-08 instance, the operating system image used is Ubuntu 20.04 LTS. The specific versions of the software and libraries used in our system are as follows:
MySQL database version: 8.0.33;
HElib version: 2.2.2;
g++ version: 9.4.0;
Boost C++ library version: 1.71.
Furthermore, some important dependencies are as follows.
Number theory library (NTL) version: 11.4.3;
Multiple-precision arithmetic library (GMP) version: 6.2.0.
All the source code files are compiled with the following flags and options.
\textit{-fPIC}: This flag generates position-independent code, which is required for loadable functions.
\textit{-fopenmp}: This flag enables OpenMP support for thread-level parallelization.
\textit{-std=C++17}: This option sets the C++ language standard to C++17.
These specific configurations and versions ensure compatibility and enable the desired features and functionality in our system implementation.

\subsubsection{Baseline Systems}

Rache~\cite{otawose_sigmod23} was introduced as a technique to accelerate partial homomorphic encryption (PHE) schemes, such as Paillier and Symmetria. The fundamental concept behind Rache is to decompose an arbitrary plaintext into a summation of additive terms, each of which can be retrieved from a caching pool. In this work, we choose the radix as two, meaning that any given plaintext is expressed as the sum of power-of-two terms. By exploiting the commutativity of addition, this process can be parallelized across multiple CPU cores, resulting in significant speedup.
However, Rache has two main limitations:
(i) It can only be applied to additive homomorphic encryption schemes, as it relies on the properties of addition.
(ii) It is applicable solely to integer values and does not support floating-point numbers.
Furthermore, the randomization of ciphertexts in Rache incurs some computational cost. This is because the randomness is accumulated across all possible radixes that are cached in memory, adding to the overall processing overhead.

\begin{table*}[th]
  \begin{center}
    \caption{List of benchmarks}
    \label{tbl:benchmark}
    \begin{tabular}{l c c c c c c}
\toprule
    Benchmark & \# of Records & Data Type & Minimal Value & Maximal Value & Average Value & Standard Variance\\
\midrule
    Covid19 & 341 & Integer & 123,021 & 2,309,884 & 1,063,465.029 & 570,009.089\\
    Bitcoin & 1,086 & Float & 274,252.698 & 9,999,999.999 & 7,412,197.895 & 3,472,109.034\\
    hg38 & 34,424 & Integer & 1 & 360 & 10.915 & 9.891\\
    P\_Size & 200,000 & Integer & 1 & 50 & 25.427 & 14.441\\
    P\_RetailPrice & 200,000 & Float & 901.00 & 2,098.99 & 1,499.49 & 294.673\\
    O\_TotalPrice & 1,500,000 & Float & 857.71 & 555,285.16 & 151,219.537 & 88,621.401\\
    L\_ExtendedPrice & 6,001,215 & Float & 901.00 & 104,949.50 & 38,255.138 & 23,300.436\\
\bottomrule
    \end{tabular}
  \end{center}
\end{table*}

BGV~\cite{bgv} is a full homomorphic encryption (FHE) scheme implemented in HElib. It is considered a second-generation FHE scheme as it departs from the original ideal-lattice-based approach for reducing ciphertext noise. Instead, BGV relies on the hardness of the learning with error (LWE) problem and incorporates optimizations like re-linearization and modulus-switching.
BGV is based on the ring-variant of learning with error (RLWE), which posits that solving a system of polynomial rings with small noises is computationally difficult. One key advantage of BGV is its packed representation in ciphertext, allowing for the encryption of a vector of plaintext messages (e.g., 24 plaintexts in our experiment) within the same ciphertext. This improves both speed and storage efficiency.
However, BGV has two notable limitations:
(i) It is most efficient when applied to integers. While applying BGV to floating-point numbers is possible by expanding the plaintext representation, controlling the noise becomes challenging, and the cost of reducing the noise becomes impractical.
(ii) BGV consumes significant memory space due to the sizable ciphertexts.
In our evaluation, we utilize the following parameters for BGV.
Plaintext prime modulus: p = 4,999;
Cyclotomic polynomial argument: m = 32,109;
Number of bits in the modulus chain: bits = 500;
Number of columns in the key-switching matrix: c = 2.
It is important to note that one BGV ciphertext can correspond to multiple plaintexts (e.g., 24). However, MySQL loadable functions only support record-level encryption, which poses limitations on the application of BGV in this context.

CKKS~\cite{ckks} is an FHE scheme that supports complex numbers as well as real numbers, making it particularly useful for applications involving privacy-preserving machine learning. It shares several design principles with BGV, including leveled noise reduction, reliance on the hardness of the RLWE problem, and the use of packed ciphertext to represent plaintext vectors.
In this work, we employ CKKS with the following parameters.
Cyclotomic index: m = $2^{14}$;
Ciphertext modulus: bits = 119 bits;
Precision: precision = 20 bits;
Key-switching matrices: c = 2 columns.
Similar to BGV, CKKS is designed to encrypt a large number of plaintext "slots" (e.g., $2^{12}$) that are packed into a single ciphertext to reduce the amortized per-plaintext cost. 
However, implementing this packing, also known as "batching," in MySQL loadable functions poses challenges because MySQL only supports interfaces for single records or aggregation of entire groups in SQL statements.

\subsubsection{Workloads}

We collect seven data sets to evaluate this work.
Four of them are from the standard TPC-H benchmark and three of them are from various real-world applications.
Key characteristics of these workloads are summarized in Table~\ref{tbl:benchmark}.

The primary benchmark for this work is TPC-H ver. 3.0.1~\cite{tpch3}.
TPC-H allows the user to specify the scales of the generated data;
in this paper we set the scale as one, resulting in about one gigabyte of data.
We will focus on four attributes from three tables:
\textit{Part.P\_Size}, \textit{Part.P\_RetailPrice}, \textit{Orders.O\_TotalPrice}, and \textit{Lineitem.L\_ExtendedPrice}.
The number of tuples ranges between 200,000 and more than six million.
The first real-world application is the U.S. national COVID-19 statistics from April 2020 to March 2021~\cite{covid19data}.
The data set has 341 days of 16 metrics, such as \textit{death increase}, \textit{positive increase}, and \textit{hospitalized increase}.
The second real-world application is the history of Bitcoin trade volume~\cite{bitcoin_trade} since it was first exchanged in the public in February 2013.
The data consists of the accumulated Bitcoin exchange on a 3-day basis from February 2013 to January 2022,
totaling 1,086 floating-point numbers.
The third real-world application is the human genome reference 38~\cite{hg_data},
commonly known as \textit{hg38},
which includes 34,424 rows of singular attributes,
e.g., \textit{transcription positions}, \textit{coding regions}, and \textit{number of exons}, last updated in March 2020.

\subsection{Computational Cost of Different Homomorphic Functionalities}

This section reports the performance of vanilla CKKS when it is integrated into MySQL loadable functions.
We are particularly interested in three functionalities (i.e., algorithms):
multiplication between a plaintext and a ciphertext \textit{EvalMulPlain},
addition between two ciphertexts \textit{EvalAdd},
and the encryption of a plaintext \textit{Enc}.

Table~\ref{tbl:func_ckks} reports the performance of the three important functionalities of CKKS in MySQL loadable functions.
We repeat each of the three algorithms 1,000 times over random floating-point numbers between 1 and 20,000 and report the average.
The variances of these repeated experiments are negligible and therefore are not reported.

\begin{table}[h]
  \begin{center}
    \caption{Comparing the computational cost of CKKS~\cite{ckks} algorithms (i.e., functionalities) in MySQL loadable functions}
    \label{tbl:func_ckks}
    \begin{tabular}{l r c}
\toprule
    CKKS Functionality & Raw Time & Relative Speed \\
\midrule
    \textit{CKKS\_Enc} & 2.93 ms & 1$\times$ \\
    \textit{CKKS\_EvalAdd} & 0.17 ms & 17$\times$ \\
    \textit{CKKS\_EvalMulPlain} & 0.08 $\mu$s & 36,625$\times$ \\
\bottomrule
    \end{tabular}
  \end{center}
\end{table}

We can learn from Table~\ref{tbl:func_ckks} that 
(i) the addition between two ciphertexts in CKKS is more than one order of magnitude faster than the encryption algorithm and
(ii) the plaintext-ciphertext multiplication in CKKS is more than four orders of magnitude faster than the encryption algorithm.
We stress that those speedups are regarding the computational cost only and should not be directly interpreted into an estimate of the end-to-end encryption protocol, 
such as Rache or Streche.
This is because the encryption protocol involves other costs such as aggregation operation on top of \textit{EvalAdd}/\textit{EvalMulPlain} and the memory manipulation in the database buffer pool.

Table~\ref{tbl:func_bgv} reports the corresponding functionalities implemented in BGV~\cite{bgv}.
We see that this FHE scheme designed for integers exhibits a larger computational discrepancy:
the addition operation among ciphertexts is more than two orders of magnitude (which is on par with the numbers reported in~\cite{otawose_sigmod23});
the plaintext multiplication is five orders of magnitude faster than the encryption.

\begin{table}[h]
  \begin{center}
    \caption{Comparing the computational cost of key BGV~\cite{bgv} algorithms (i.e., functionalities) in MySQL loadable functions}
    \label{tbl:func_bgv}
    \begin{tabular}{l r c}
\toprule
    BGV Functionality & Raw Time & Relative Speed \\
\midrule
    \textit{BGV\_Enc} & 50.26 ms & 1$\times$ \\
    \textit{BGV\_EvalAdd} & 0.22 ms & 228$\times$ \\
    \textit{BGV\_EvalMulPlain} & 0.18 $\mu$s & 279,222$\times$ \\
\bottomrule
    \end{tabular}
  \end{center}
\end{table}

\subsection{Caching Overhead}

This section reports the caching overhead during the offline stage when a variety of concurrent CPU threads are invoked.
While MySQL does not implement a multi-threading mechanism for a specific query,
in Streche we manually employ OpenMP~\cite{openmp} to activate multiple threads to accelerate the precomputing of cached ciphertexts.
In addition to the caching time,
we also record the memory consumption for different numbers of threads.
The number of threads is dynamically set at runtime through the 
\textit{omp\_set\_num\_threads()}
function.

Table~\ref{tbl:openmp_ckks} reports the time and memory consumption when the offline caching of 200,000 plaintexts is carried out by 1 -- 96 threads.
Recall that our test bed is equipped with dual CPUs and the overall number of threads is 96.
However, the overly fine granularity of splitting the caching task beyond 32 threads contributes less speedup than the cost of context switching among threads.
When 32 threads are involved, the offline stage takes about 32 seconds to precompute 200,000 ciphertexts for online encryption tasks.
In terms of memory consumption,
the cost is pretty consistent (81 GB) except when all of the CPU cores are taken (113 GB).
However, this is usually not a concern in practice because the full scale of CPU cores should not be taken due to the suboptimal time cost (i.e., 137.46 vs. 31.81 seconds).

\begin{table}[h]
  \begin{center}
    \caption{Comparing the cost of different numbers of threads for caching ciphertexts in Streche-CKKS and Streche-BGV}
    \label{tbl:openmp_ckks}
    \begin{tabular}{l | r r | c c}
\toprule
    Number of & \multicolumn{2}{c|}{Running Time (second)} & \multicolumn{2}{c}{Memory (GB)} \\
     Threads & \multicolumn{1}{c}{CKKS} & BGV & CKKS & BGV\\
\midrule
    1 & 650.90 & 1,893.88 & 65 & 146 \\
    2 & 337.87 & 956.68 & 65 & 146 \\
    4 & 183.04 & 490.87 & 65 & 146\\
    8 & 99.24 & 290.60 & 65 & 146 \\
    16 & 62.42 & 152.98 & 65 & 146 \\
    32 & 31.81 & 89.28 & 65 & 146 \\
    64 & 62.60 & 103.95 & 65 & 146 \\
    96 & 137.46 & OOM & 107 & 192+ \\
\bottomrule
    \end{tabular}
  \end{center}
\end{table}

Table~\ref{tbl:openmp_ckks} also reports Streche's performance regarding thread parallelism of precomputing 50,000 plaintexts based on the BGV~\cite{bgv} scheme.
We observe a larger overhead in terms of both running time and memory consumption,
which indicates that BGV is not as efficient as CKKS (note that the workload in Streche-BGV is only a quarter of that in Streche-CKKS).
However, it is worth noting that both BGV and CKKS incur minimal offline overhead when 32 threads are invoked. 
We did not report the numbers for 96 threads because the experiment ran out of memory (OOM), i.e., memory consumption exceeds 192 GB.
Due to the large memory overhead incurred by BGV and its limitation for floating-point numbers,
in the remainder of this section, we will focus on CKKS-based Streche.

\subsection{Additive Streche for Integers}

We evaluate the effectiveness of Streche when it is applied to additively homomorphic encryption schemes.
The baseline of this experiment is the state-of-the-art caching mechanism,
namely Rache~\cite{otawose_sigmod23},
which applies only to integers.
Therefore, for some benchmarks evaluated in this section,
the original values are truncated from floating-point numbers into integers.
As discussed before, we will focus on CKKS;
although CKKS is originally designed for complex numbers,
it can surely suffice to encrypt integers.

Table~\ref{tbl:streche_a} reports the computational time of AStreche and compares it with that of Rache~\cite{otawose_sigmod23}.
The large variety of benchmarks leads to a broad spectrum of running time ranging between sub-seconds and hours.
However, we observe that the speedup of AStreche over Rache is significant and somewhat consistent,
i.e., between 2.3$\times$ (Covid19) and 3.3$\times$ (Bitcoin).

\begin{table}[h]
\begin{threeparttable}
    \caption{Encrypting integers with static caching (Rache~\cite{otawose_sigmod23}) and streamed caching (this work) in MySQL}
    \label{tbl:streche_a}
    \begin{tabular}{l r r c}
\toprule
    Benchmark & \multicolumn{1}{c}{Rache} & \multicolumn{1}{c}{This Work} & Speedup\\
\midrule
    Covid19 & 907 ms & 387 ms & 2.3$\times$ \\
    Bitcoin & 4.2 sec & 1.3 sec & 3.3$\times$ \\
    hg38 & 49 sec & 17 sec & 2.9$\times$\\
    P\_Size & 195 sec & 68 sec & 2.9$\times$ \\
    P\_RetailPrice & 369 sec & 131 sec & 2.8$\times$ \\
    O\_TotalPrice & 60 min & 21 min & 2.8$\times$ \\
    L\_ExtendedPrice & 218 min & 76 min & 2.9$\times$ \\
\bottomrule
    \end{tabular}
\end{threeparttable}  
\end{table}

\subsection{Full Streche for Floating-point Numbers}

FStreche leverages the efficient constant multiplication among the elements in the ciphertext space.
Since there does not exist prior work on streamed caching for FHE schemes\footnote{Recall that Rache~\cite{otawose_sigmod23} only supports additive operations on ciphertexts.},
the only baseline system to compare with is the original FHE,
in our case, the CKKS scheme~\cite{ckks}.

Table~\ref{tbl:streche_f} reports the performance of FStreche and compares it to that of vanilla CKKS~\cite{ckks}.
Again, the effectiveness of FStreche is evident:
the speedup ranges between 1.8$\times$ and 5.6$\times$.
Careful readers would notice that the range is wider than that of AStreche (cf.~Table~\ref{tbl:streche_a}).
We stress that the numbers in both tables are not comparable in general because the experiments conducted for AStreche are carried out on the truncated integers if the benchmark is over original floating-point numbers (e.g., P\_RetailPrice, O\_TotalPrice, L\_ExtendedPrice).

\begin{table}[h]
  \begin{center}
    \caption{Performance of floating-point numbers using CKKS~\cite{ckks} and the streamed caching (this work) in MySQL}
    \label{tbl:streche_f}
    \begin{tabular}{l r r c}
\toprule
    Benchmark & \multicolumn{1}{c}{CKKS} & \multicolumn{1}{c}{This Work} & Speedup\\
\midrule
    Covid19 & 931 ms & 287 ms & 3.2$\times$ \\
    Bitcoin & 2.9 sec & 1.6 sec & 1.8$\times$ \\
    hg38 & 92 sec & 16 sec & 5.6$\times$\\
    P\_Size & 528 sec & 94 sec & 5.6$\times$ \\
    P\_RetailPrice & 532 sec & 215 sec & 2.5$\times$ \\
    O\_TotalPrice & 62 min & 36 min & 1.8$\times$ \\
    L\_ExtendedPrice & 265 min & 137 min & 1.9$\times$ \\
\bottomrule
    \end{tabular}
  \end{center}
\end{table}

\section{Additional Related Work}

The optimization of performance for Homomorphic Encryption (HE) schemes has been extensively studied in existing literature. For instance, researchers have explored hardware-based optimization techniques \cite{nsam_micro21,dreis_vlsi20,ydoroz_tc15} for HE schemes. A recent article highlights the memory wall as the current performance bottleneck of HE \cite{decastro2021does}. However, our work is distinct from these studies as it focuses on purely algorithmic and software optimization of the $ASEnc$ and $FSEnc$ methods, employing a carefully chosen caching mechanism.

The concept of \textit{incremental cryptography} was first formally introduced in the 1990s \cite{mbellare_crypto94,mbellare_stoc95}, primarily from a theoretical standpoint. Recent work on incremental encryption schemes can be found in \cite{imironov_eurocrypt12,pananth_eurocrypt17,lkhati_sac18}. Incremental encryption has gained significant attention in current research, particularly for efficient data encoding in resource-constrained contexts like mobile computing \cite{fwang_fgcs21,gke_journal21,tbhatia_ccpe20}. However, to the best of our knowledge, no existing cryptosystem simultaneously supports both homomorphic encryption and incremental encoding while ensuring proven semantic security.

\section{Conclusion}

This paper presents two novel encryption algorithms aimed at accelerating homomorphic encryption in cloud-based outsourced databases. The first algorithm, called $ASEnc$, enables efficient caching of integer polynomials with low-cost randomization, surpassing the performance of the state-of-the-art caching scheme Rache~\cite{otawose_sigmod23}. The second algorithm, named $FSEnc$, focuses on streaming the cached floating-point numbers to enhance fully homomorphic encryption schemes. We provide theoretical proof of the correctness and semantic security of both algorithms.
To validate our proposals, we implement the algorithms as loadable functions within MySQL, hosted on the Chameleon Cloud~\cite{katek_atc20}. We conduct extensive evaluations using seven benchmark tests. The experimental results demonstrate that $ASEnc$ achieves a performance improvement of 2.3--3.3$\times$ compared to Rache, while $FSEnc$ surpasses the state-of-the-art floating-point fully homomorphic encryption scheme CKKS~\cite{ckks} by 1.8--5.6$\times$.
All source code, datasets, dependent libraries, log files, and experimental results used in this study are accessible through a shared image titled `Streche' through the \textit{cc-snapshot} utility on the Chameleon Cloud.

\textit{Results presented in this paper were obtained using the Chameleon testbed supported by the National Science Foundation.}

\bibliographystyle{plain}
\bibliography{ref_new}

\begin{thebibliography}{10}

\bibitem{aalmamun_sc21}
Abdullah Al-Mamun, Feng Yan, and Dongfang Zhao.
\newblock {BAASH}: Lightweight, efficient, and reliable blockchain-as-a-service
  for hpc systems.
\newblock In {\em International Conference on High Performance Computing,
  Networking, Storage and Analysis (SC)}, 2021.

\bibitem{pananth_eurocrypt17}
Prabhanjan Ananth, Aloni Cohen, and Abhishek Jain.
\newblock Cryptography with updates.
\newblock In Jean-S{\'e}bastien Coron and Jesper~Buus Nielsen, editors, {\em
  Advances in Cryptology -- EUROCRYPT 2017}, pages 445--472, Cham, 2017.
  Springer International Publishing.

\bibitem{dbeaver_crypto91}
Donald Beaver.
\newblock Efficient multiparty protocols using circuit randomization.
\newblock In Joan Feigenbaum, editor, {\em Advances in Cryptology - {CRYPTO}
  '91, 11th Annual International Cryptology Conference, Santa Barbara,
  California, USA, August 11-15, 1991, Proceedings}, volume 576 of {\em Lecture
  Notes in Computer Science}, pages 420--432. Springer, 1991.

\bibitem{mbellare_crypto94}
Mihir Bellare, Oded Goldreich, and Shafi Goldwasser.
\newblock Incremental cryptography: The case of hashing and signing.
\newblock In Yvo Desmedt, editor, {\em Advances in Cryptology - {CRYPTO} '94,
  14th Annual International Cryptology Conference, Santa Barbara, California,
  USA, August 21-25, 1994, Proceedings}, volume 839 of {\em Lecture Notes in
  Computer Science}, pages 216--233. Springer, 1994.

\bibitem{mbellare_stoc95}
Mihir Bellare, Oded Goldreich, and Shafi Goldwasser.
\newblock Incremental cryptography and application to virus protection.
\newblock In Frank~Thomson Leighton and Allan Borodin, editors, {\em
  Proceedings of the Twenty-Seventh Annual {ACM} Symposium on Theory of
  Computing (STOC)}, 1995.

\bibitem{tbhatia_ccpe20}
Tarunpreet Bhatia, A.K. Verma, and Gaurav Sharma.
\newblock Towards a secure incremental proxy re-encryption for e-healthcare
  data sharing in mobile cloud computing.
\newblock {\em Concurrency and Computation: Practice and Experience (CCPE)},
  32(5):e5520, 2020.
\newblock e5520 CPE-18-0794.R1.

\bibitem{bitcoin_trade}
{Bitcoin Trade History}.
\newblock \url{https://www.blockchain.com/charts/trade-volume}, Accessed 2022.

\bibitem{bgv}
Zvika Brakerski, Craig Gentry, and Vinod Vaikuntanathan.
\newblock (leveled) fully homomorphic encryption without bootstrapping.
\newblock In {\em Proceedings of the 3rd Innovations in Theoretical Computer
  Science Conference}, ITCS '12, page 309–325, New York, NY, USA, 2012.
  Association for Computing Machinery.

\bibitem{bv11}
Zvika Brakerski and Vinod Vaikuntanathan.
\newblock Efficient fully homomorphic encryption from (standard) lwe.
\newblock In {\em 2011 IEEE 52nd Annual Symposium on Foundations of Computer
  Science}, pages 97--106, 2011.

\bibitem{ckks}
Jung~Hee Cheon, Andrey Kim, Miran Kim, and Yong~Soo Song.
\newblock Homomorphic encryption for arithmetic of approximate numbers.
\newblock In Tsuyoshi Takagi and Thomas Peyrin, editors, {\em Advances in
  Cryptology - {ASIACRYPT} 2017 - 23rd International Conference on the Theory
  and Applications of Cryptology and Information Security, Hong Kong, China,
  December 3-7, 2017, Proceedings, Part {I}}, volume 10624 of {\em Lecture
  Notes in Computer Science}, pages 409--437. Springer, 2017.

\bibitem{covid19data}
{Covid-19 Data}.
\newblock \url{ https://covidtracking.com/data/download/national-history.csv},
  Accessed 2022.

\bibitem{mdaum_icde21}
M.~Daum, B.~Haynes, D.~He, A.~Mazumdar, and M.~Balazinska.
\newblock Tasm: A tile-based storage manager for video analytics.
\newblock In {\em 2021 IEEE 37th International Conference on Data Engineering
  (ICDE)}, pages 1775--1786, Los Alamitos, CA, USA, apr 2021. IEEE Computer
  Society.

\bibitem{decastro2021does}
Leo de~Castro, Rashmi Agrawal, Rabia Yazicigil, Anantha Chandrakasan, Vinod
  Vaikuntanathan, Chiraag Juvekar, and Ajay Joshi.
\newblock Does fully homomorphic encryption need compute acceleration?, 2021.

\bibitem{ydoroz_tc15}
Yarkın Doroz, Erdinc Ozturk, and Berk Sunar.
\newblock Accelerating fully homomorphic encryption in hardware.
\newblock {\em IEEE Transactions on Computers}, 64(6):1509--1521, 2015.

\bibitem{elgamal_tit85}
T.~Elgamal.
\newblock A public key cryptosystem and a signature scheme based on discrete
  logarithms.
\newblock {\em IEEE Transactions on Information Theory}, 31(4):469--472, 1985.

\bibitem{bfv}
Junfeng Fan and Frederik Vercauteren.
\newblock Somewhat practical fully homomorphic encryption.
\newblock Cryptology ePrint Archive, Paper 2012/144, 2012.
\newblock \url{https://eprint.iacr.org/2012/144}.

\bibitem{fraleigh_book03}
J.B. Fraleigh and V.J. Katz.
\newblock {\em A First Course in Abstract Algebra}.
\newblock Addison-Wesley series in mathematics. Addison-Wesley, 2003.

\bibitem{cgentry_stoc09}
Craig Gentry.
\newblock Fully homomorphic encryption using ideal lattices.
\newblock In {\em Proceedings of the Forty-first Annual ACM Symposium on Theory
  of Computing (STOC)}, 2009.

\bibitem{bhaynes_sigmod21}
Brandon Haynes, Maureen Daum, Dong He, Amrita Mazumdar, Magdalena Balazinska,
  Alvin Cheung, and Luis Ceze.
\newblock Vss: A storage system for video analytics.
\newblock In {\em Proceedings of the 2021 International Conference on
  Management of Data}, SIGMOD/PODS '21, page 685–696, New York, NY, USA,
  2021. Association for Computing Machinery.

\bibitem{helib}
{HElib}.
\newblock \url{https://github.com/homenc/HElib}, Accessed 2022.

\bibitem{ntru}
Jeffrey Hoffstein, Jill Pipher, and Joseph~H. Silverman.
\newblock {NTRU:} {A} ring-based public key cryptosystem.
\newblock In Joe Buhler, editor, {\em Algorithmic Number Theory, Third
  International Symposium, ANTS-III, Portland, Oregon, USA, June 21-25, 1998,
  Proceedings}, volume 1423 of {\em Lecture Notes in Computer Science}, pages
  267--288. Springer, 1998.

\bibitem{wkuan_clsr18}
W.~Kuan Hon and Christopher Millard.
\newblock Banking in the cloud: Part 3 – contractual issues.
\newblock {\em Computer Law \& Security Review}, 34(3):595--614, 2018.

\bibitem{hg_data}
{Human Genome Databases}.
\newblock \url{http://hgdownload.soe.ucsc.edu/goldenPath/hg38/database/},
  Accessed 2022.

\bibitem{tkanwal_cluster21}
Tehsin Kanwal, Adeel Anjum, and Abid Khan.
\newblock Privacy preservation in e-health cloud: taxonomy, privacy
  requirements, feasibility analysis, and opportunities.
\newblock {\em Clust. Comput.}, 24(1):293--317, 2021.

\bibitem{gke_journal21}
Gang Ke, Shi Wang, and Huan-huan Wu.
\newblock Parallel incremental attribute-based encryption for mobile cloud data
  storage and sharing.
\newblock {\em Journal of Ambient Intelligence and Humanized Computing}, pages
  1--11, 01 2021.

\bibitem{katek_atc20}
Kate Keahey, Jason Anderson, Zhuo Zhen, Pierre Riteau, Paul Ruth, Dan
  Stanzione, Mert Cevik, Jacob Colleran, Haryadi~S. Gunawi, Cody Hammock, Joe
  Mambretti, Alexander Barnes, Fran\c{c}ois Halbach, Alex Rocha, and Joe
  Stubbs.
\newblock Lessons learned from the chameleon testbed.
\newblock In {\em Proceedings of the 2020 USENIX Annual Technical Conference
  (USENIX ATC '20)}. USENIX Association, July 2020.

\bibitem{lkhati_sac18}
Louiza Khati and Damien Vergnaud.
\newblock Analysis and improvement of an authentication scheme in incremental
  cryptography.
\newblock In Carlos Cid and Michael J.~Jacobson Jr., editors, {\em Selected
  Areas in Cryptography - {SAC} 2018 - 25th International Conference, Calgary,
  AB, Canada, August 15-17, 2018, Revised Selected Papers}, volume 11349 of
  {\em Lecture Notes in Computer Science}, pages 50--70. Springer, 2018.

\bibitem{ecc}
Neal Koblitz.
\newblock Elliptic curve cryptosystems.
\newblock {\em Mathematics of Computation}, 48(177):203--209, January 1987.

\bibitem{imironov_eurocrypt12}
Ilya Mironov, Omkant Pandey, Omer Reingold, and Gil Segev.
\newblock Incremental deterministic public-key encryption.
\newblock In David Pointcheval and Thomas Johansson, editors, {\em Advances in
  Cryptology -- EUROCRYPT 2012}, pages 628--644, Berlin, Heidelberg, 2012.
  Springer Berlin Heidelberg.

\bibitem{aes}
{National Institute and Technology of Standards}.
\newblock Advanced encryption standard.
\newblock {\em NIST FIPS PUB 197}, 2001.

\bibitem{openmp}
{OpenMP}.
\newblock \url{http://openmp.org}, Accessed May 2023.

\bibitem{ppail_eurocrypt99}
Pascal Paillier.
\newblock Public-key cryptosystems based on composite degree residuosity
  classes.
\newblock In {\em Proceedings of the 17th International Conference on Theory
  and Application of Cryptographic Techniques}, EUROCRYPT'99, page 223–238,
  Berlin, Heidelberg, 1999. Springer-Verlag.

\bibitem{dreis_vlsi20}
Dayane Reis, Jonathan Takeshita, Taeho Jung, Michael Niemier, and Xiaobo~Sharon
  Hu.
\newblock Computing-in-memory for performance and energy-efficient homomorphic
  encryption.
\newblock {\em IEEE Transactions on Very Large Scale Integration (VLSI)
  Systems}, 28(11):2300--2313, 2020.

\bibitem{rsa}
R.~L. Rivest, A.~Shamir, and L.~Adleman.
\newblock A method for obtaining digital signatures and public-key
  cryptosystems.
\newblock {\em Commun. ACM}, 21(2):120–126, feb 1978.

\bibitem{nsam_micro21}
Nikola Samardzic, Axel Feldmann, Aleksandar Krastev, Srinivas Devadas, Ronald
  Dreslinski, Christopher Peikert, and Daniel Sanchez.
\newblock {\em F1: A Fast and Programmable Accelerator for Fully Homomorphic
  Encryption}, page 238–252.
\newblock Association for Computing Machinery, 2021.

\bibitem{symmetria_vldb20}
Savvas Savvides, Darshika Khandelwal, and Patrick Eugster.
\newblock Efficient confidentiality-preserving data analytics over
  symmetrically encrypted datasets.
\newblock {\em Proc. VLDB Endow.}, 13(8):1290–1303, April 2020.

\bibitem{sealcrypto}
{M}icrosoft {SEAL} (release 3.7).
\newblock \url{https://github.com/Microsoft/SEAL}, September 2021.
\newblock Microsoft Research, Redmond, WA.

\bibitem{otawose_sigmod23}
Olamide T.~Tawose, Jun Dai, Lei Yang, and Dongfang Zhao.
\newblock Toward efficient homomorphic encryption for outsourced databases
  through parallel caching.
\newblock {\em Proceedings of the ACM on Management of Data (SIGMOD)}, May
  2023.

\bibitem{tpch3}
{TPC-H 3.0.0}.
\newblock \url{
  http://tpc.org/tpc_documents_current_versions/current_specifications5.asp},
  Accessed May 2023.

\bibitem{fwang_fgcs21}
Fenghe Wang, Junquan Wang, and Wenfeng Yang.
\newblock Efficient incremental authentication for the updated data in fog
  computing.
\newblock {\em Future Generation Computer Systems (FGCS)}, 114:130--137, 2021.

\bibitem{xwang_aaai20}
Xinying Wang, Olamide~Timothy Tawose, Feng Yan, and Dongfang Zhao.
\newblock {HDK:} toward high-performance deep-learning-based kirchhoff
  analysis.
\newblock In {\em The Thirty-Fourth {AAAI} Conference on Artificial
  Intelligence (AAAI)}, 2020.

\bibitem{xzhu_tdsc21}
Xiaojie Zhu, Erman Ayday, Roman Vitenberg, and Narasimha~Raghavan Veeraragavan.
\newblock Privacy-preserving search for a similar genomic makeup in the cloud.
\newblock {\em IEEE Transactions on Dependable and Secure Computing}, 2021.

\end{thebibliography}

\end{document}